\documentclass[%
  reprint,
  superscriptaddress,
showpacs,preprintnumbers,
nofootinbib,
 amsmath,amssymb, 
 aps,
 prx
floatfix,
]{revtex4-2}
\usepackage{amsmath,amstext,amssymb}
\usepackage[T1]{fontenc}
\usepackage{aas_macros}
\usepackage{xcolor}
\usepackage{soul}
\usepackage{comment}
\usepackage{graphicx}
\usepackage[normalem]{ulem}
\usepackage[export]{adjustbox}
\usepackage[colorlinks=true,allcolors=blue]{hyperref}
\usepackage{orcidlink}
\usepackage{multirow}
\newcommand{\cosvar}{\Vec{\theta}}

\begin{document}

\title[Standard Sirens with BBHs in AGNs]{Standard Siren Cosmology with
Gravitational Waves from Binary Black Hole
Mergers in Active Galaxy Nuclei}

\author{C. R. Bom~\orcidlink{0000-0003-4383-2969}}\email{debom@cbpf.br}
\affiliation{Centro Brasileiro de Pesquisas F\'isicas, Rua Dr. Xavier Sigaud 150, 22290-180 Rio de Janeiro, RJ, Brazil}
\affiliation{Centro Federal de Educa\c{c}\~{a}o Tecnol\'{o}gica Celso Suckow da Fonseca, Rodovia M\'{a}rcio Covas, lote J2, quadra J - Itagua\'{i} (Brazil)}
\author{A.~Palmese~\orcidlink{https://orcid.org/0000-0002-6011-0530}}\email{palmese@cmu.edu}
\affiliation{McWilliams Center for Cosmology, Department of Physics, Carnegie Mellon University, Pittsburgh, PA 15213, USA}



\begin{abstract}
The detection of gravitational waves (GW) with an electromagnetic counterpart enabled the first Hubble Constant $H_0$ measurement through the standard siren method. Current constraints suggest that $\sim 20-30\%$ of LIGO/Virgo/KAGRA (LVK) Binary Black Hole (BBH) mergers might occur in Active Galactic Nuclei (AGN) disks. The claim for a possible association of several BBH mergers with flaring AGNs suggests that cosmological analyses using BBH and AGNs might be promising. We explore standard siren analyses through a method that takes into account the presence of background flaring AGNs, without requiring a unique host galaxy identification, and apply it to realistic GW simulations. Depending on the fraction of LVK BBHs that induce flares, we expect to constrain $H_0$ at the $\sim 6-7\%$ ($\sim 4-5\%$) precision with $\sim 2-3$ years or $\sim 160-240$ events ($\sim 1$ year or $500$ events) of LVK at design (A+) sensitivity, assuming that systematic BBH follow-up searches are performed. 
We also show that in a scenario where only $\sim 1\%$ of the BBHs induce detectable flares it is possible to achieve an $H_0$ precision from $7.5\%$ to $15\%$ with $\sim 240$ events. In addition, a $\sim 5-27\%$ precision is achievable with complete AGN catalogs and 1 year of LVK run,  without the need of any flare identification.
\end{abstract}




\maketitle

\section{Introduction}\label{sec1}
The detection of gravitational Waves (GW) by LIGO and Virgo has enabled a variety of novel measurements of cosmological parameters. The ``standard siren'' method, first proposed by \cite{schutz}, relies on a luminosity distance measurement from the GW detection, and together with an independent measurement of the redshift of the GW event, it allows us to probe the expansion of the Universe through the distance-redshift relation.
The current $4-6\sigma$ tension between the measurement of the Hubble constant $H_0$ from Cosmic Microwave Background \cite{planckcolab} and Cepheid-anchored Type Ia Supernova \cite{riess2021} analyses makes an independent, standard siren measurement of $H_0$ from gravitational waves a promising endeavor to shed light on the origin of this tension through an independent probe. However, despite enormous efforts by the astronomical community, only one optical counterpart to a GW event, GW170817 \cite{ligobns,MMApaper}, has been confidently identified, and the identification of its host galaxy along with its redshift enabled standard siren measurements of $H_0$ \citep{firststandardsiren,2023arXiv230519914P,Hotokezaka}. About $\sim 50$ similar associations are needed to produce a measurement at the precision level needed to discern between the two $H_0$ measurements currently in major discrepancy \cite{2019PhRvL.122f1105F}. 
In the cases where no counterpart is identified it is possible to use a different approach, the ``dark siren'' method, by taking advantage of assumptions on the mass distribution of compact objects (e.g. \cite{Farr_2019,spectral_sirens}), or by taking into account the ensemble of potential host galaxies \cite{schutz,darksiren1,Hitchhiker}. 
Recent dark siren analyses \cite{LVK21_StS,darksiren_DESI,finke2021cosmology,palmese20_sts,2023RNAAS...7..250B,2023RNAAS...7..250B} reach a precision on $H_0$ down to $\sim 20\%$. Hundreds to thousands of such events (or a few ``special'' events \cite{Borhanian}) will be required to reach a 2\% uncertainty on $H_0$ (e.g. \cite{diaz22}). 
For current generation GW detectors, most of the dark siren works make use of binary black hole (BBH) mergers. BBHs are usually expected to originate as isolated binaries (e.g. \cite{Belczynski_2002}) or through dynamical channels (e.g. \cite{Rodriguez_2016,McKernan12,Yang_2019,Conselice_20,palmese_conselice, Yang_agn_2019}). A promising site for the formation of massive stellar mass binaries through hierarchical mergers is in Active Galactic Nuclei (AGN) \cite{McKernan12}. Recent works \cite{ford_mcKernan_21_rate,Gayathri_2021,Gayathri2023} show that $\sim 20\%$ of the BBHs detected by LIGO/Virgo are expected to occur in AGNs. 
In the AGN environment, the electromagnetic signature associated with BBHs is contingent upon the interplay between the merger and the adjacent disk gas. Generally, a BBH within an AGN disk is displaced from the point of merger, exhibiting a recoil kick velocity determined by the mass asymmetry and spin orientation of the progenitor system. The gas in the disk around the progenitor BBH tends to follow the merged remnant, yet it encounters the neighboring AGN disk gas, consequently generating a luminous shockwave within the disk. Gas accretion onto the kicked black hole may give rise to a Bondi tail, or a jet may be launched and produce an afterglow. These phenomena may be observable in a wide range of electromagnetic radiation wavelengths, including in the optical/ultraviolet \cite{BartosRapid,mckernanRam,kimura2021,tagawa23,rodriguez2023optical}. 

Several AGN flares have been flagged as possible electromagnetic counterparts \cite{graham20,graham23}, but given the variable nature of AGNs, it is challenging to establish a confident association with low probability of chance coincidence for GW events with a broad sky localization, and multiple flares may occur within the GW region around the same time. On the other hand, one could consider using all of the AGNs that exist within the GW localization, in case no follow-up observations exist or a flare is not expected (e.g. for obscured AGNs). Even though it is challenging to identify which flare or which AGN is indeed associated with a single GW event, after enough observations of GW events along with AGN flares or AGN catalogs, it will be possible, statistically, to say whether the GW-AGN associations are consistent with a background of chance coincidence observations or they are indeed related, given that the spatial and temporal distributions of chance associations will be different from that of a real association \cite{Bartos_2017,Palmese_AGN,veronesi}. If the association is deemed significant, then the AGN population also provides the redshift information needed for standard siren measurements.

We explore whether making use of the population of BBH occurring in AGNs can improve standard siren constraints, considering that AGNs are a small subsample of the generic galaxy population such as that used in dark siren analyses so far, and that the number of host galaxies one marginalizes over dictates the final cosmological parameter precision. 
Furthermore, the distance reach of BBH mergers (luminosity distance $\sim 1400-1600$ Mpc for the LIGO detectors in the current observing run \cite{Abbott2020_prospects}), as opposed to that of binary neutron star (BNS) mergers ($\sim 160-190$ Mpc), will allow us to eventually probe cosmological parameters beyond $H_0$, such as the Universe matter density and the dark energy equation of state. Note that this approach is different from that of previous work assuming a unique host galaxy is identified with high confidence \cite{yang,Mukherjee2020,Gayathri21,Chen_2022} through identification of a unique, confirmed electromagnetic counterpart. Motivated by the findings of \cite{graham23} and the expected sky localizations of GW events in the coming years, it is reasonable to assume that at least for current generation GW detectors high confidence associations on single events will be challenging. We base our forecast on the assumption that the associations can only be made at a population level, and that is also the way our standard siren approach is implemented. Because of this, our constraints will be very different from those from previous works. First, precision: because the uncertainty in counterpart association is taken into account here, we expect our forecasts to be less stringent and more realistic than a forecast assuming a high confidence association. Second, accuracy: our simulations include background flares as potential counterparts, yet our method yields unbiased constraints, as we show in the following sections. On the other hand, if one uses the classical bright standard siren approach with a flare that is unlikely to be associated, it is clear that the analysis may yield biased constraints, since one would use an incorrect value of the redshift.

In this work, we explore the promising avenue of taking advantage of the BBH events that merge in AGN disks to constrain cosmological parameters. The formalism was first presented in \cite{Palmese_AGN}, where the authors show how to constrain the fraction of LIGO/Virgo BBH that give rise to AGN flares, while also deriving cosmological parameters. The method is generic and can be used to constrain the fraction of BBHs hosted by AGNs (as also explored in \cite{Bartos_2017}) and cosmological parameters without the requirement of a flare, hence of follow-up observations, assuming that a complete AGN catalog is available. 

\section{Methods}\label{sec:methods}

\subsection{Standard Siren Cosmology with Noisy Source
Identification Environment}

The association between BBHs and AGN flares might be addressed as a signal-to-background distinction, as done for the formalism presented in \cite{Palmese_AGN}. The total number of AGN flares per solid angle $\Omega$ and redshift $z$, $\frac{dN}{d\Omega dz}$, can be construed as a composite model consisting of an AGN flare associated to the gravitational wave (GW) event, if any, and the background number density of AGN flares within a specified temporal interval. $\frac{dN}{d\Omega dz}$ is dependent on $\lambda$, the fraction of BBHs inducing AGN flares. For a set of events, we can then write down a posterior to constrain $\lambda$. 

The galaxy catalog approach of the dark siren method uses the redshift information of possible host galaxies of the GW event and the luminosity distance obtained from GW detectors. Differently from a traditional dark siren approach, we consider AGN flares within the high probability sky location of the GW event occurring within a time window from the BBH event detection, to constrain the fraction of events that induce a ``signal'' flare, i.e. $\lambda$. Note that by including this factor and letting it be $\lambda<1$, we are allowing for the fact that not all BBH, but only a fraction, may occur in AGN. Similarly, one can consider known AGNs (instead of flaring AGNs or all possible host galaxies) where a follow-up is not available, in which case we refer to $\lambda_{\rm AGN}$. Thus, the proposed method takes advantage of an intermediate perspective between the dark and bright siren approaches considered so far.

 Let $N$ be the number of GW events with GW data $\left\{x_i^\mathrm{GW}\right\}_{i = 1}^N \equiv \left\{\Omega_{i}^\mathrm{GW}, d_{L,i}^\mathrm{GW}\right\}_{i =1}^N$ and with AGN data $\left\{x_i^\mathrm{AGN}\right\}_{i = 1}^N \equiv \left\{\left\{\Omega_{ij}^\mathrm{AGN}, z_{ij}^\mathrm{AGN}\right\}_{j=1}^{k}\right\}_{i =1}^N$, where $\Omega$ is a sky line of sight, $d_L$ is the luminosity distance and $z$ is the redshift. The subscript $i$ refers to a GW event, while $j$ refers to an AGN potentially associated to the event $i$. After applying Bayes' theorem and marginalizing over the fraction of GW events hosted by AGNs $\lambda$, the posterior on the cosmological parameters $\cosvar$ is given by:
\begin{align}
    p \left( \cosvar \text{\textbar} \left\{x_i^{\mathrm{AGN}} \right\}_{i = 1}^N ), 
    \left\{x^\mathrm{GW}_i\right\}_{i = 1}^N \right) 
   & \propto p(\cosvar) \int d\lambda p(\lambda) \prod_i^N \mathcal{L}_i \left( \cosvar, \lambda \right) 
\end{align}
The likelihood of observing the GW data $x_i^\mathrm{GW}$ and $k$ AGNs or AGN flares with sky positions and redshifts $\left\{\Omega_{ij}^\mathrm{AGN}, z_{ij}^\mathrm{AGN}\right\}_{j=1}^{k}$, is given by an inhomogeneous Poisson process. We then marginalize over the  position of the GW source $(\Omega_i^\mathrm{GW}, z_i^\mathrm{GW})$, to obtain the likelihood $\mathcal{L}_i\left(\cosvar, \lambda \right)$ as:
\begin{align}
    & \mathcal{L}_i\left(\cosvar, \lambda \right) \propto  \nonumber \\
    &\propto \prod_{j = 1}^k \big[ \lambda p(x_i^\mathrm{GW} \mid \Omega_{ij}^\mathrm{AGN}, d_L(z_{ij}^\mathrm{AGN}, \cosvar))~p_0(\Omega_{ij}^\mathrm{AGN}, z_{ij}^\mathrm{AGN}) \nonumber \\
    & + R_B(\Omega_{ij}^\mathrm{AGN}, z_{ij}^\mathrm{AGN}, \cosvar ) \big] e^{-\mu_i}.  \label{eq:Li}
\end{align}
In Eq. \ref{eq:Li} the term that multiplies $\lambda$ is effectively the GW posterior on the sky location and redshift evaluated at the AGN position and redshift, for a given cosmology. The second term in brackets is the background rate, and it is defined as $R_B=T \frac{dB}{d\Omega dz dt}$, i.e. the rate of background flares or AGNs per unit solid angle and redshift within a time period $T$. At last, the term $\mu_i$ is the expected number of observed AGN flares or AGNs:
\begin{equation}
    \mu_i \equiv \int \frac{dN_i}{d\Omega dz} P_\mathrm{det}^\mathrm{AGN}(\Omega,z) d\Omega dz\, ,\label{eq:selection}
\end{equation}
where $P_\mathrm{det}^\mathrm{AGN}$ is the detection probability of the AGNs/flares. For a full derivation of this likelihood see \cite{Palmese_AGN}. We sample the posterior on the cosmological parameters and $\lambda$ using the python package \texttt{emcee}. 

 We use realistic simulations of GW events for LIGO/Virgo/KAGRA (LVK) close to the expected design sensitivity (and fourth observing run, O4, sensitivity) and for their A+ sensitivity, along with their expected 3D localization reconstructions, produced as described in Section \ref{sec:gwsims}. To estimate the number of AGNs within a GW localization volume, we use a fiducial number of $n_{\rm AGN}=10^{-4.75}$ Mpc$^{-3}$\cite{Greene2007,Greene2009,2017NatCo...8..831B}, and also consider a second scenario with higher background contamination ($n_{\rm AGN}=10^{-4}$ Mpc$^{-3}$, motivated by the fact that BBH merger may extend to low accretion rates AGNs \cite{Yang_2019}). The number of background AGNs flaring depends on several factors. As most of AGNs present some variability, the definition  might depend on the rise in brightness timescale $\tau$, and also be sensitive to the variability model chosen. For instance, the first flare claimed to be of BBH origin in AGN J124942.3+344929 presented a  a rise of $\Delta m\sim 0.4$ mag over  $\sim 50$ days \cite{graham20}. For this event, \cite{Palmese_AGN} used structure functions (SF) to derive a probability of flare $\alpha_{\rm flare} \sim 10^{-4}$ with $\Delta m >0.4$. On the other hand \cite{graham20} estimated, $\alpha_{\rm flare}=5 \times 10^{-6}$ considering a fit in the sample of AGNs data from the Zwicky Transient Facility (ZTF). As the number of background flares can be influenced by the exact theoretical modeling, which is yet to be well understood for BBH events, and the timescales probed by a specific follow-up campaign, we choose the more conservative estimation from \cite{Palmese_AGN} (i.e. higher background flare contamination than that estimated in e.g. \cite{graham20}) scenario, one similar to the AGN J124942.3+344929 flare background estimate, $\alpha_{\rm flare}=10^{-4}$.

Even though \cite{ford_mcKernan_21_rate} find that up to $\sim 80\%$ of BBHs detected by LIGO/Virgo/KAGRA might originate from the AGN channel, many could be hosted in an accretion disk obscured from view (type-II AGNs). Although flares occurring in this type of AGN might still become detectable at later times and at longer, 
IR wavelengths (see, for instance \cite{10.1093/mnras/staa2351}), this may impact the fraction of BBHs inducing flares, hence the $\lambda$ 
we consider in the case of flaring AGNs should be lower than that of \cite{ford_mcKernan_21_rate}, which will include some obscured AGNs. 
Using the prescription from \cite{Hopkins_2007} the fraction of unobscured AGNs varies with the bolometric luminosity, it can be $\sim 20\%$ ($\sim 26\%$) for AGNs of luminosity $10^{44}$($10^{46}$) erg s$^{-1}$ visible at optical wavelengths. In the case of GW190521 in association with AGN J124942.3$+$344929, considering AGNs with $g<20.5$, the type-I AGNs comprise $\sim 22\%$ of the entire AGN sample \cite{Palmese_AGN}. Thus, if only BBHs flares from type-I AGNs were detectable, this would limit $\lambda$ from $<0.8$ to 
$<0.18$. Therefore, in our fiducial scenario, we restrict the analysis to $\lambda \in [0.05,0.3]$, where the lower limit is chosen by selecting type-I AGNs 
for $\sim 20\%$ of BBH in AGNs found by \cite{Gayathri2023}. Furthermore, since \cite{Gayathri_2021} find a lower limit for the AGN channel of $\sim 3\%$ of all BBH events, we also consider an alternative, more pessimistic scenario, where $\lambda$ can go down to 1/4 of the lower bound in our fiducial scenario ($\lambda \in [0.0125,0.025 ]$). We note that the lower bound here is not $\sim 22\%$ of the 3\% value from \cite{Gayathri_2021}, as bringing $\lambda$ below 1\% would not produce any interesting results, at least for the upcoming gravitational wave runs.

\subsection{Gravitational Wave simulations \label{sec:gwsims}}

\begin{figure}[h]%
\centering
\includegraphics[width=0.49\textwidth]{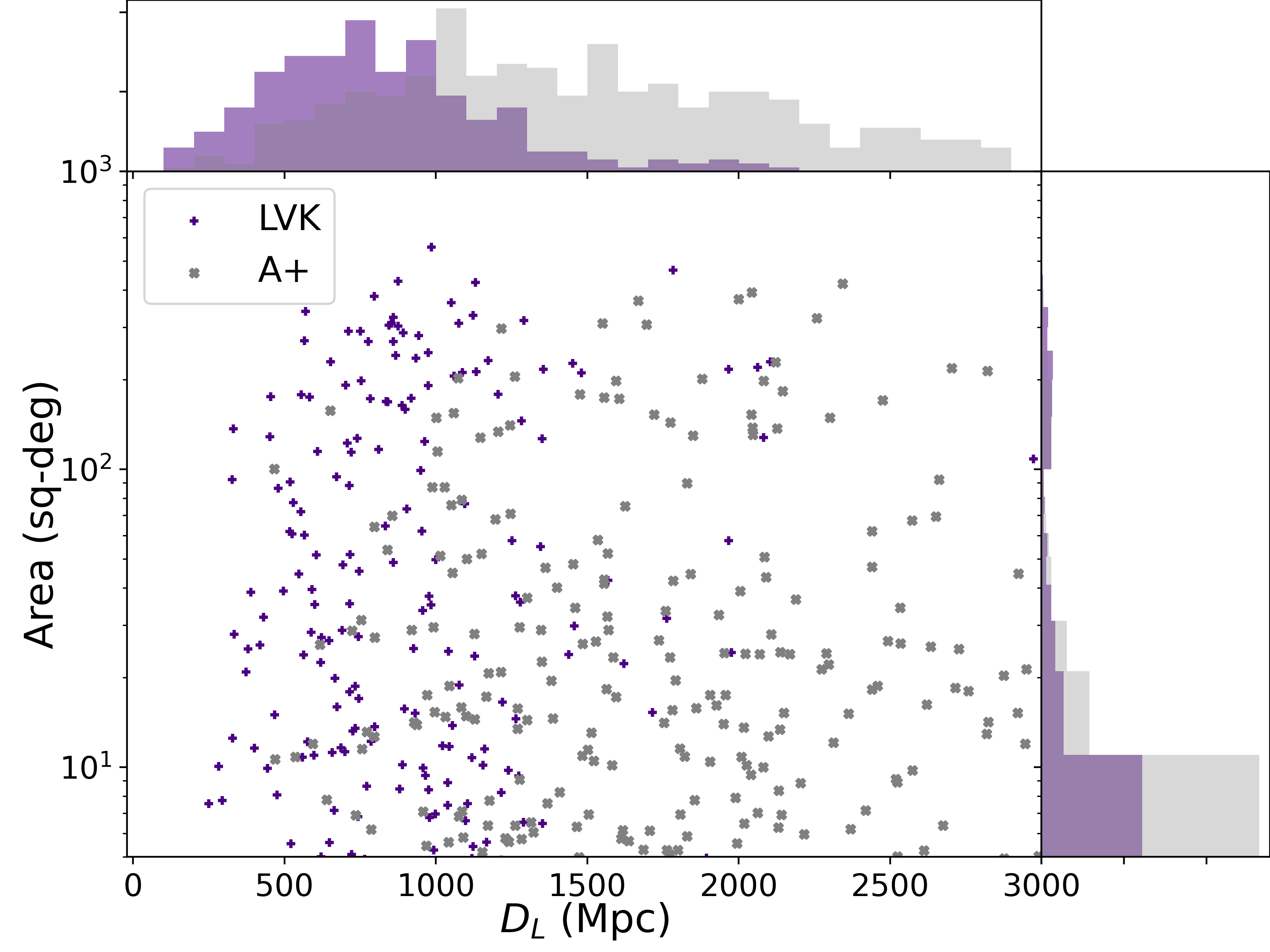}
\includegraphics[width=0.49\textwidth]{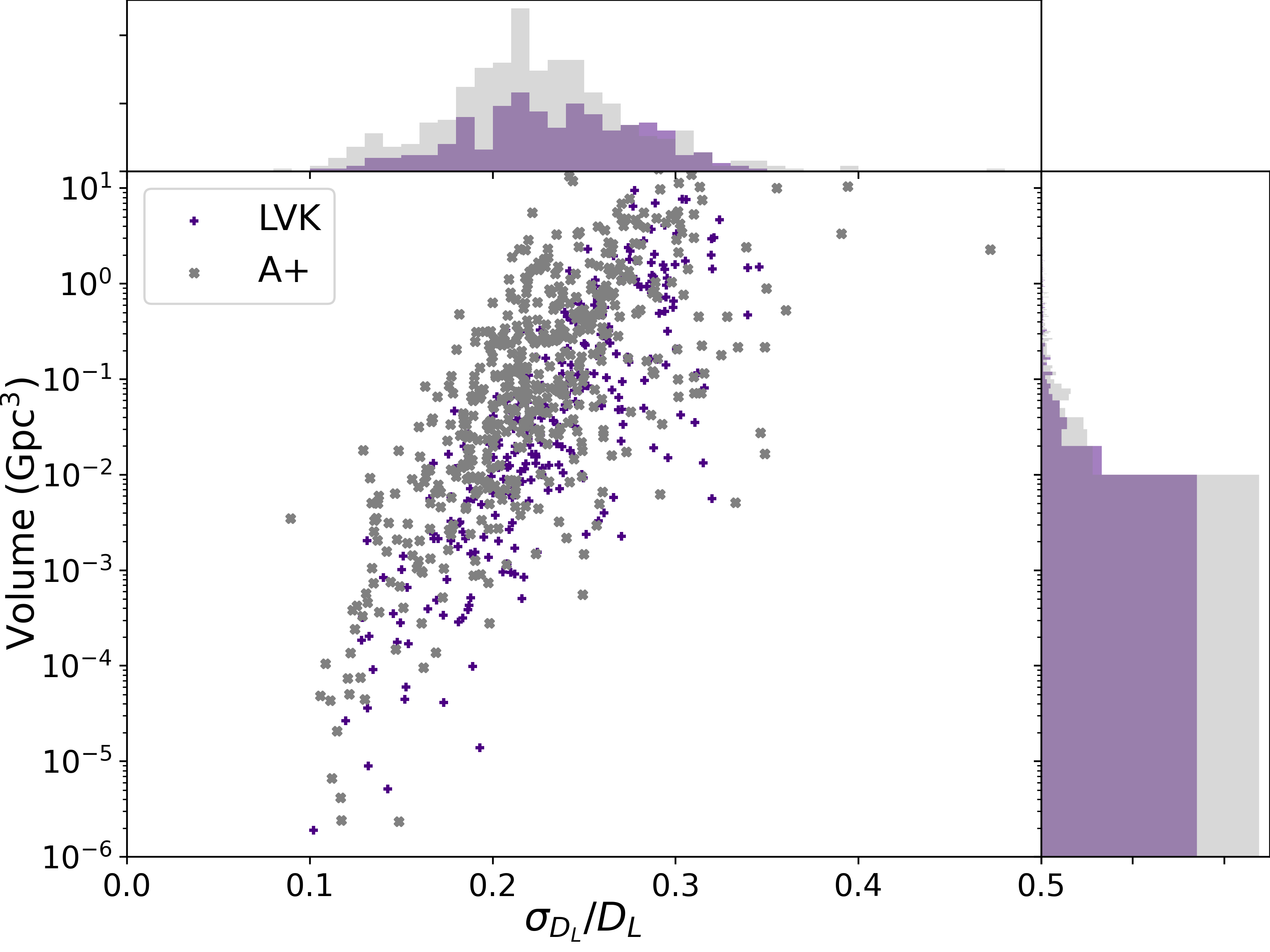}
\caption{The LVK (O4, indigo) and LVK A+ (O5, gray) parameters set obtained from simulations of binary black hole mergers. Top: distance and  $90\%$ credible interval sky localization area of our simulated events. Bottom: The GW $90\%$ credible interval Volume and the fractional error in the GW Luminosity distance.}\label{fig:sims}
\end{figure}

We analyze a dataset of simulated BBH mergers generated according to the expected observations from the current LIGO-Virgo-KAGRA (LVK) observing run (the fourth run, ``O4''), and the following LVK A+ upgrade (potentially close to what expected for the fifth observing run, O5). Our methodology is similar to that employed in \cite{Palmese_AGN,petrov2022data, bom2023designing}. Gravitational wave events are generated using the \texttt{BAYESTAR}~\cite{Singer_2016,bayestar} software, which makes use of tools from \texttt{LALSuite} \cite{lalsuite}. We adopt sensitivity curves expected for Advanced LIGO and Virgo at design sensitivity and for KAGRA at a BNS inspiral range of 80 Mpc, and at A+ sensitivity, all taken from \url{https://dcc.ligo.org/LIGO-T2000012-v1/public}, considering a duty cycle of $70\%$ \cite{Abbott2020_prospects}. In the A+ scenario, we also include a LIGO-India detector. The BBH population follows a mass distribution in the form of a power law plus peak with parameters from \cite{LVK:2021duu}. We set the input cosmological parameters based on a flat $\Lambda$CDM cosmology with parameters from \cite{planckcolab}. In Figure \ref{fig:sims} we show the 90\% CI area versus luminosity distance (top panel), and the comoving volume versus fractional distance error (bottom panel) for the simulations used in this work. Clearly, with A+ we will be able to detect more events and up to larger distances than with LVK at design sensitivity, resulting in a larger fraction of high comoving volume events. At the same time, with A+ a large number of extremely well localized 10 sq. deg. events will be available. 

We use the events' 90\% comoving volume to compute the expectation value of the number of background flares or AGNs for each event, and draw for each simulated event a random number of background objects following a Poisson distribution. The background objects follow a uniform in comoving volume redshift distribution, and an isotropic distribution in the sky area (i.e. we ignore AGN clustering for the purpose of this work). On the other hand, whether a GW event has an associated ``signal'' flare, depends on the input value of $\lambda$, again following a Poisson distribution. Signal flares follow a different sky and redshift distribution, which is dictated by the  GW event posteriors.

\section{Results}\label{sec2}

\subsection{AGN flares case}

\begin{figure}[h]%
\centering
\includegraphics[width=0.49\textwidth]{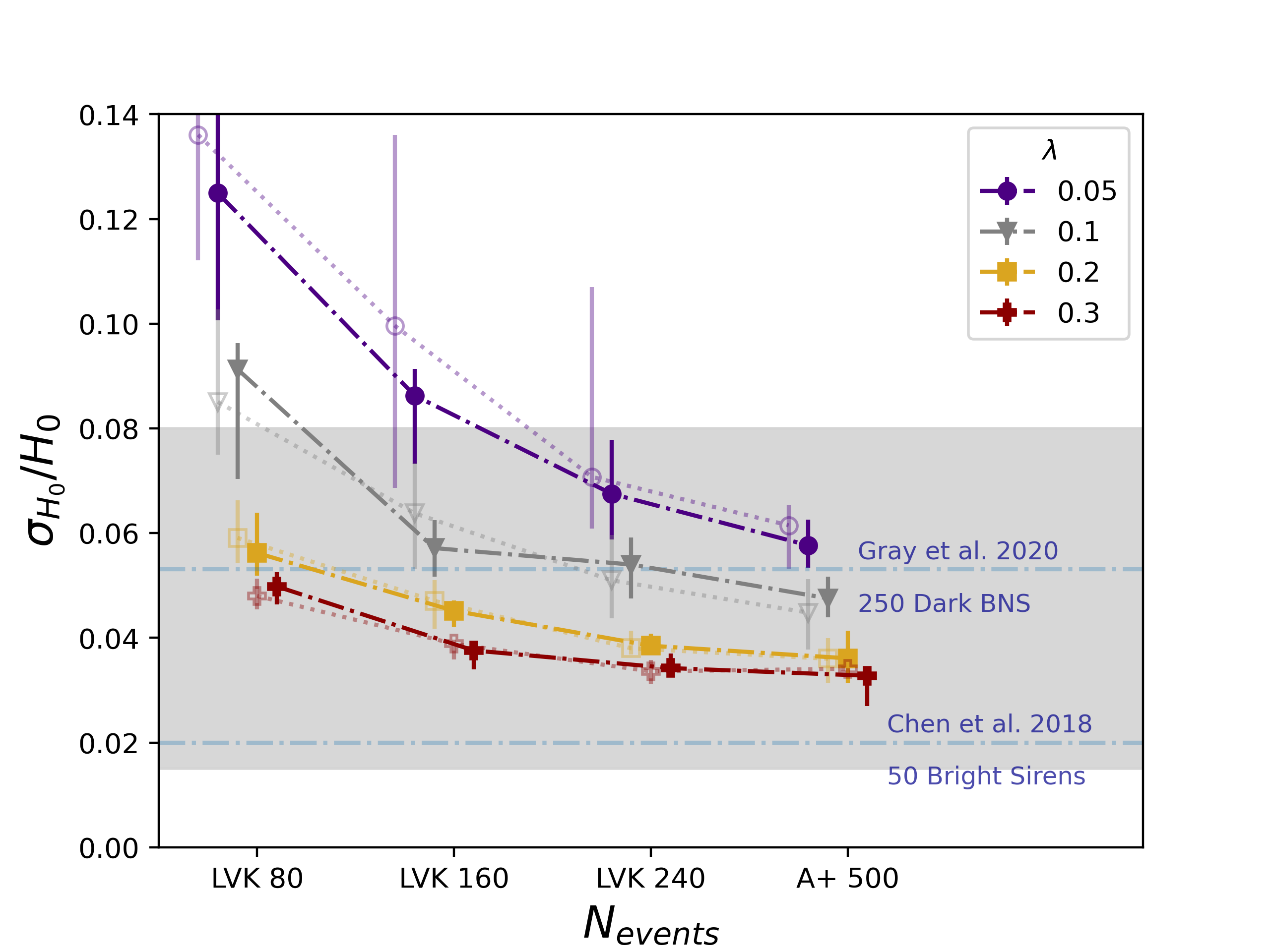}
\includegraphics[width=0.49\textwidth]{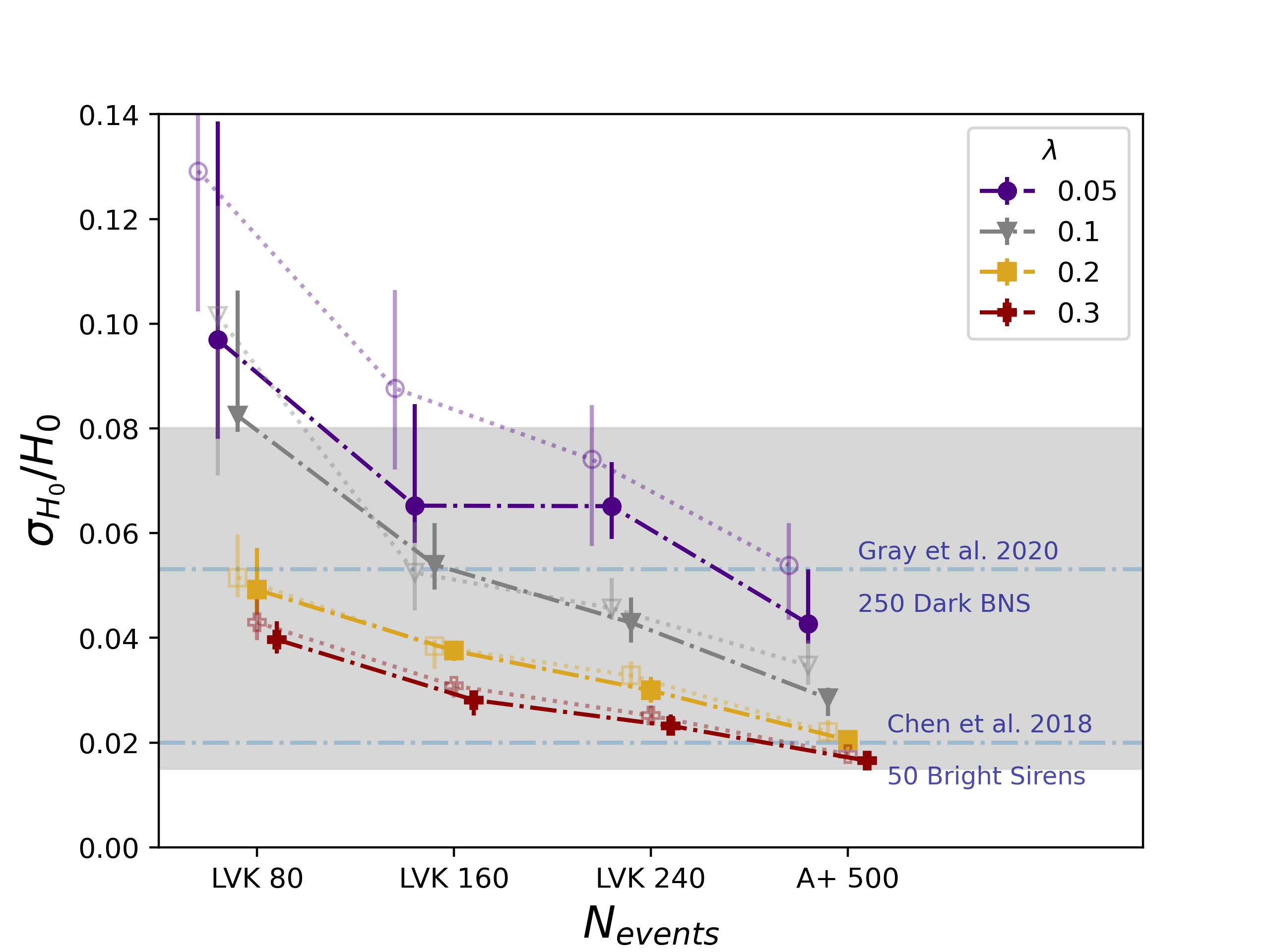}
\caption{Hubble constant precision as a function of number of events from LVK (similarly to an O4 design sensitivity) and LVK A+ (similar to O5) for different input values of the fraction of BBHs inducing AGN flares $\lambda$. The upper dot-dashed horizontal line presents the expected $H_0$ precision from $250$ dark sirens using BNSs from \cite{gray2020}, considering a $50\%$ complete galaxy catalog and unweighted galaxies. The lower dot-dashed line presents the forecast from \cite{chen17} after $\sim 50$ BNS bright sirens. The shaded region shows the uncertainties over $100$ realizations expected for a golden dark siren where we can identify the host from HLV+ with spectroscopic redshift  \cite{Borhanian}. The top (bottom) panel considers a flat prior over $\Omega_m \in [0,1.0]$ ($[0.25,0.35]$). The number density of AGN is $n_{\rm AGN}=10^{-4.75}$  Mpc$^{-3}$ for the filled symbols and $10^{-4}$ Mpc$^{-3}$ for the empty symbols.\label{fig:H0_precision}}
\end{figure}

\begin{figure}[h]%
\centering
\includegraphics[width=0.49\textwidth]{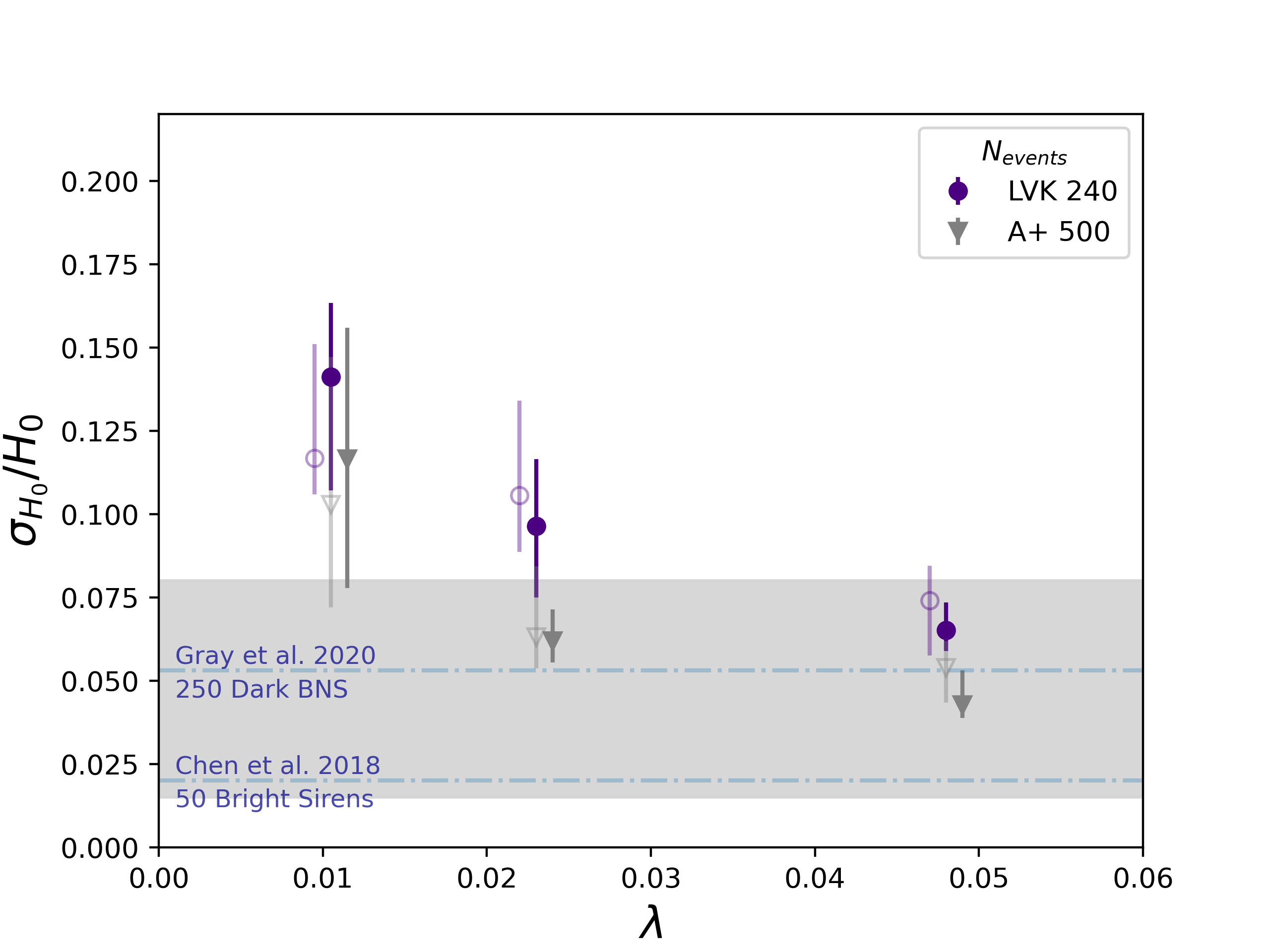}
\caption{Hubble constant precision as a function of $\lambda$ from LVK design and LVK A+ for the most pessimistic scenarios considered here, with a lower limit on the fraction of BBH producing flaring AGNs of $\sim 10^{-2}$. The upper dot-dashed horizontal line presents the expected $H_0$ precision from $250$ Dark Siren using BNSs from \cite{gray2020}, considering a $50\%$ complete galaxy catalog and unweighted galaxies. The lower dot-dashed line presents the forecast from \cite{chen17} after $\sim 50$ BNS bright sirens. The shaded region shows the uncertainties over $100$ realizations expected for a golden dark siren where we can identify the host from HLV+ with spectroscopic redshift. The number density AGN is $n_{\rm AGN}=10^{-4.75}$  Mpc$^{-3}$ for the filled symbols and $10^{-4}$ Mpc$^{-3}$ for the empty symbols. These results consider a flat prior on $\Omega_m \in [0.25,0.35]$.\label{fig:H0_precision2}}
\end{figure}

\begin{figure}[h]%
\centering
\includegraphics[width=0.49\textwidth]{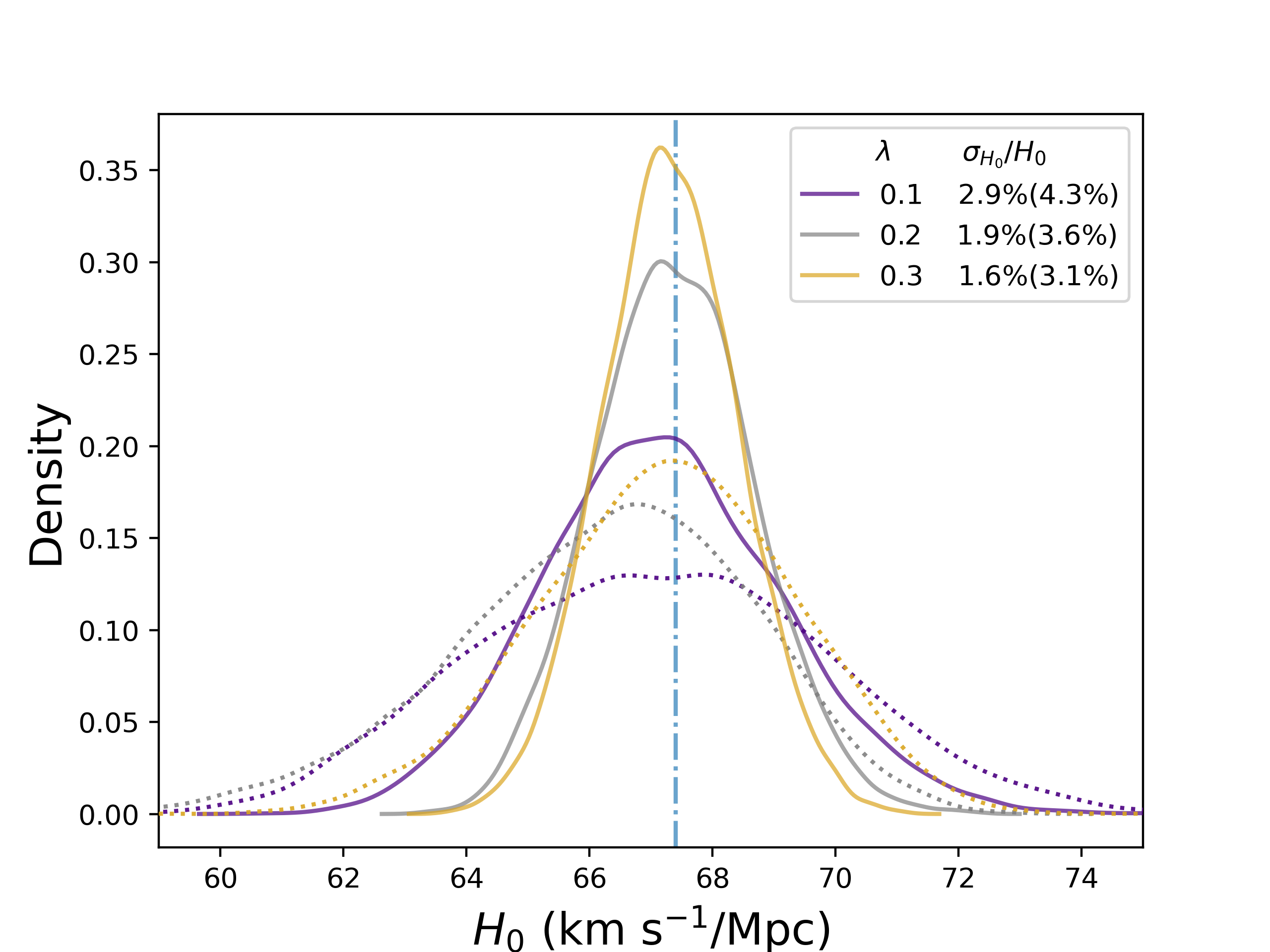}
\includegraphics[width=0.49\textwidth]{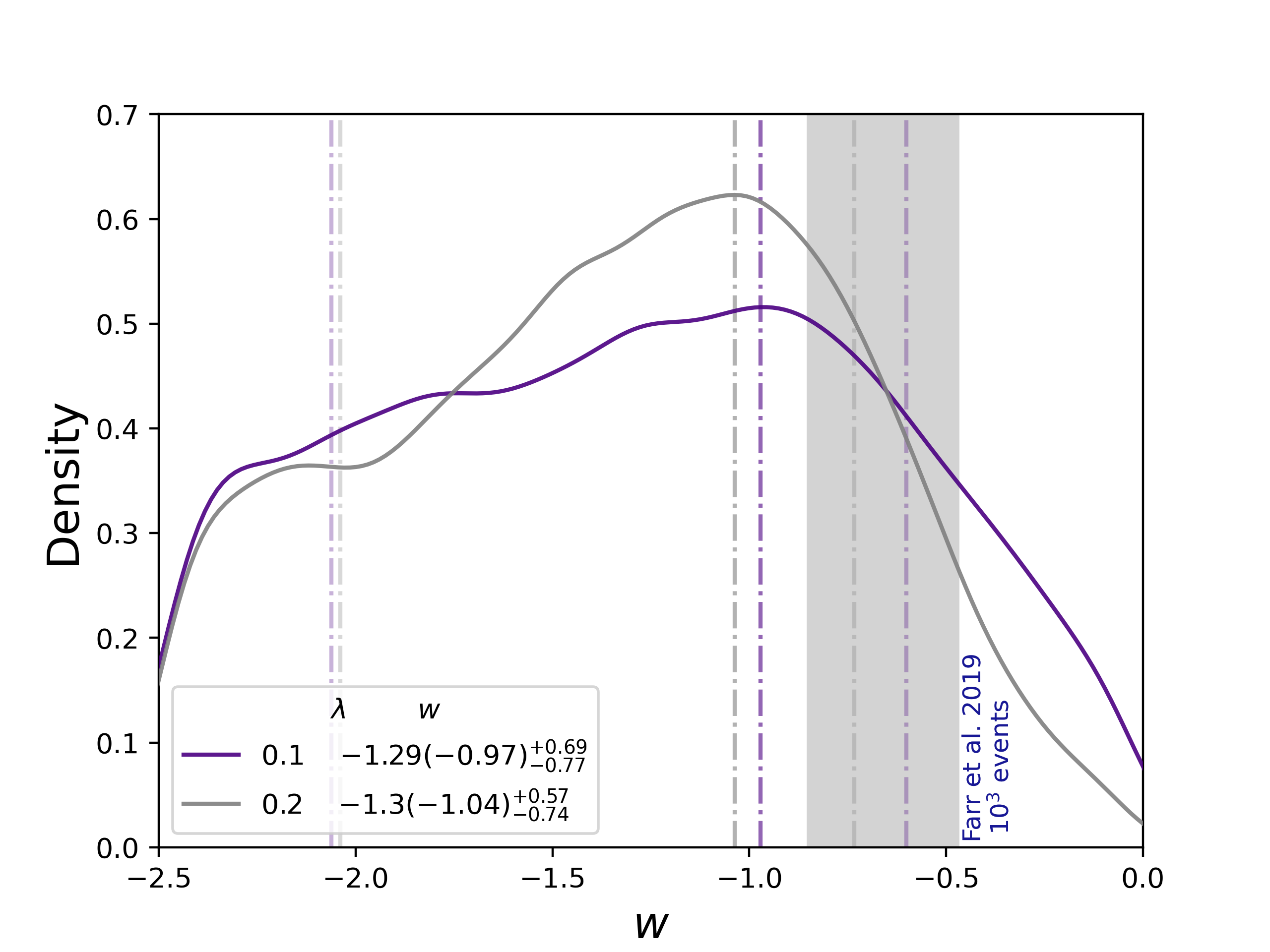}
\caption{Posteriors from $1$ year (or $500$ events) of LVK A+ within a flaring AGN scenario with $n_{\rm AGN}=10^{-4.75}$ Mpc$^{-3}$ and $\alpha_{\rm flare}=10^{-4}$. \emph{Top:} $H_0$  posterior considering a $\Lambda$CDM model for different input values of $\lambda$. The solid lines represent the result for a restricted flat prior of $\Omega_m \in [0.25,0.35]$ with the precision quoted in the legend. The dashed lines assume a flat wide prior $\Omega_m \in [0.0,1.0]$, and their respective recovered precision is quoted in parentheses in the legend. The vertical line is the input value of $H_0$. \emph{Bottom:} The dark energy equation of state parameter $w$ posterior  considering a $w$CDM cosmology for different input values of $\lambda$. The legend presents the median $w$ and the best fit (in parentheses), with the $68\%$ credible intervals. The dot-dashed lines present the $68\%$ credible intervals and the mode.  The input value of $w$ in our simulations is -1. The gray-shaded region shows the forecasts from \cite{Farr_2019} using $10^3$ LVK events. }\label{fig:H0_w}
\end{figure}

In Figure \ref{fig:H0_precision}, we present the precision on $H_0$ obtained in a flat $\Lambda$CDM cosmology with two choices of uniform priors for the matter density $\Omega_m$,  $\Omega_m \in [0.0,1.0]$ in the top panel and $\Omega_m \in [0.25,0.35]$, corresponding to  $\sim 5\sigma$ around the best fit Planck Cosmology value from \cite{planckcolab}, in the bottom panel. The prior in the Hubble constant is uniform $H_0 \in [20,140]$ km/s/Mpc, and $\lambda \in [0.0,1.0]$, as a function of the number of events in the LVK design and LVK A+ scenarios, for different input values of $\lambda$, and for the two different AGN number density considered ($n_{\rm AGN}=10^{-4.75}$ Mpc$^{-3}$ for the filled symbols, and $n_{\rm AGN}=10^{-4}$ Mpc$^{-3}$ for the empty symbols of Figure \ref{fig:H0_precision}). The points shown represent the median and the errorbars represent the $68\%$ credible interval using $10$ different runs. While the number of runs considered is on the low side for a sample that should ideally be used to derive statistics, the computational time required for a larger number of runs for all the scenarios considered would be prohibitive. 

The number of events shown corresponds to the median expectation in \cite{Abbott2020_prospects} per year in O4, i.e. $\sim 80$ events for LVK and the median expectation in \cite{petrov2022data} of $\sim 500$ events in LVK A+ (since O5 prospects are not included in the official LVK forecasts of \cite{Abbott2020_prospects}). O4 is currently scheduled for $\sim 22$ months of data acquisition, hence the median in the whole O4 run could be pushed close to $\sim 150$. Therefore, for the current observing run, the expectations might be close to what we report for 160 events. Note that \cite{petrov2022data} find a median of $106$ BBH detections in O4, and considering the uncertainties we could have up to $171$ events per year during O4. However, they use a signal-to-noise ratio (SNR) cut that is less stringent than what we use in the simulations here (which is similar to \cite{Abbott2020_prospects}), so that even if more events are expected in their case compared to \cite{Abbott2020_prospects}, they are likely to have worse localization than those we consider here, so that $\sim 120$ events for O4 is still a reasonable assumption for this work.

As expected, constraints from runs assuming a larger value of $\lambda$ tend to yield more precise measurements of the Hubble constant compared to lower $\lambda$ cases, as a larger fraction of the events considered contain the ``signal'' AGN flare, as opposed to background flares only, hence providing a meaningful contribution to the cosmology constraints. Also, as expected, the cases with a lower number density of AGNs often provide more stringent constraints than the case with larger number density. Perhaps more clearly, assuming  a flat $\Omega_m$ prior between $0.25$ and $0.35$  values provides more stringent constraints on $H_0$. In both cases, lowering $n_{\rm AGN}$ reduces the overall background count, typically resulting in better constraints on cosmological parameters,~however the improvement is often within the scatter across realizations with the higher AGN density, and appears less prominent for larger values of $\lambda$. From Figure \ref{fig:H0_precision} it is clear that if $\lambda \sim 0.1$ or greater we may reach an $H_0$ precision of $\sim 5\%$, similar to that expected from a dark siren analysis using $\sim 250$ binary neutron star mergers \cite{gray2020}, with $\sim 160$ GW events. 

 In Figure \ref{fig:H0_precision2}, we explore the more pessimistic scenarios considering values of $\lambda$ of $5\%$, $2.5\%$ and $1.25\%$. With fewer flares truly associated to BBH events compared to our fiducial case, and therefore with  a larger contamination of background events, the precision increases to $\sim 6\%-15\%$ for $\lambda =1.25-2.5\%$. This level of precision is comparable to the one available bright siren measurement  \cite{firststandardsiren}. It is fair to say that the BNS rates and the subsample of detectable bright sirens are highly uncertain and likely on the low end of previous estimates. Considering that for O4 at design sensitivity \cite{Abbott2020_prospects} we expect a median of $10$ events with a lower limit of $\sim 0$ detections and the lack of NS mergers in the first part of O4, even the scenario of $\lambda\sim 1\%$ has the potential to improve upon constraints from NS mergers with only $\sim 2$ ($\sim 5$ in O5) flares coming from BBHs out of $200$ events at LVK design sensitivity ($\sim 500$ in O5).

Next, we show posterior distributions for $H_0$ and the dark energy equation of state. The top panel of Figure \ref{fig:H0_w} presents the posteriors on $H_0$ with two choices of prior on $\Omega_m$ after one year of A+ run, the first (dashed lines) is a flat prior $\Omega_m \in [0.0,1.0]$, the second is restricted to $[0.25,0.35]$. With this choice, we find that a $\sim 3\%$ precision on $H_0$ for $\lambda=0.2$ after $160$ O4 events, or 2 years of O4. For comparison, we get $\sim 4.5\%$ precision with the broader $\Omega_m$ prior. For 500 BBH events with A+ sensitivity, we find that with the more restrictive $\Omega_m$, a 3\% precision is also possible for $\lambda=0.1$.

\begin{figure}[h]%
\centering
\includegraphics[width=0.5\textwidth]{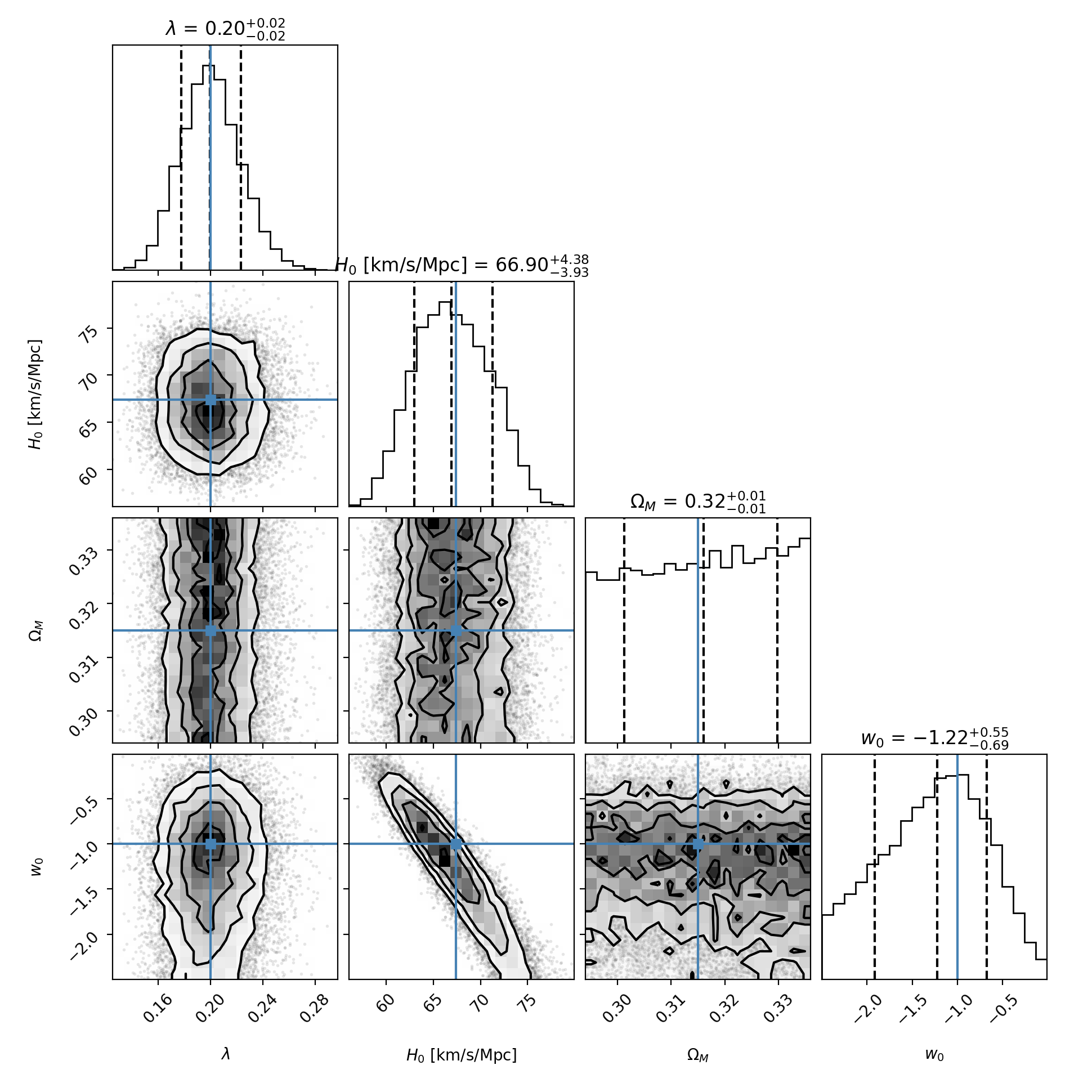}
\caption{Posteriors for one example run in a $w$CDM model with the fraction of BBH inducing flares $\lambda=0.2$, after $1$ year of LVK A+ (or 500 events) considering the AGN density $n_{\rm AGN}=10^{-4.75}$ Mpc$^{-3}$, and the fraction of AGNs flaring $\alpha_{\rm flare}=10^{-4}$. This result also considers a flat prior over $\Omega_m \in [0.25,0.35]$.}\label{fig:corner}
\end{figure}

The bottom panel of Figure \ref{fig:H0_w} shows some expected posteriors on the dark energy equation of state $w$, within a $w$CDM scenario. The results are for a one year run at A+ sensitivity. The $w$CDM model used has the same wide flat priors on $\Omega_m\in [0.0,1.0]$ and a flat prior on $w \in [-2.5,0.0]$. While the posteriors are broad, it is interesting that this method provides some constraining power in the $\Omega_m-w$ plane.   An example of a corner plot of all the parameters considered in the $w$CDM scenario is shown in Figure \ref{fig:corner}. A banana contour typical of standard candles is seen in the $\Omega_m-w$ plane, and a degeneracy is also, as expected, present between these parameters and the Hubble constant.

\begin{table*}[htp!]
    \centering

    \begin{tabular}{cc|ccccc}
    \toprule
    &&&& $\lambda_{\rm AGN}^{\rm true}$&&\\
    \hline
    BH mass  &  & $0.3$  & $0.5$ & $0.6$ & $0.7$  & $0.8$ \\
     
    \hline 
     Random & $\sigma_{H_0}/H_0 (\%)$ & $11.1^{+9.5}_{-4}$ & 
     
     $10.8^{+1.6}_{-4.0}$ & $6.2^{+1.6}_{-0.7}$ & $6.0^{+2.3}_{-0.3}$ & $5.0^{+8.8}_{-0.5}$ \\
    &$\lambda_{\rm AGN}$ & $0.34^{+0.20}_{-0.13}$ & 
    
    $0.49^{+0.23}_{-0.17}$ & $0.61^{+0.22}_{-0.18}$ & $0.70^{+0.21}_{-0.17}$ & $0.80^{+0.8}_{-0.14}$   \\
\hline 
 \\
     High-mass & $\sigma_{H_0}/H_0 (\%)$ & $26.5^{+2.0}_{-2.0}$ & 
     
     $21.2^{+5.9}_{-5.3}$ & $18.2^{+1.2}_{-5.7}$ & $15.4^{+5.5}_{-4.0}$ &
     $11.25^{+8.8}_{-1.7}$ \\
    & $\lambda_{\rm AGN}$ & $0.35^{+0.12}_{-0.03}$ & 
    
    $0.53^{+0.37}_{-0.07}$ & $0.56^{+0.40}_{-0.11}$ & $0.51^{+0.38}_{-0.06}$ & $0.48^{+0.51}_{-0.06}$   \\ \\

    \end{tabular}%
    \caption{
        \label{tab:noflare}
        Results in a scenario in which we apply the formalism to all AGNs in the field of a given GW event, i.e. not requiring a flare or follow-up campaigns. We present two cases, one in which the BBHs hosted in AGNs are randomly picked, and one in which the only occur in higher mass BBHs. For each case, the two rows show the $H_0$ precision (as median and 68\% interval from the various runs made) and the recovered median fraction of BBH mergers in AGN disks, $\lambda_{\rm AGN}$, for $80$ LVK events as a function of the input value of $\lambda_{\rm AGN}$ (given by the different columns). 
        }
\end{table*}



\begin{figure}[h]%
\centering
\includegraphics[width=0.49\textwidth]{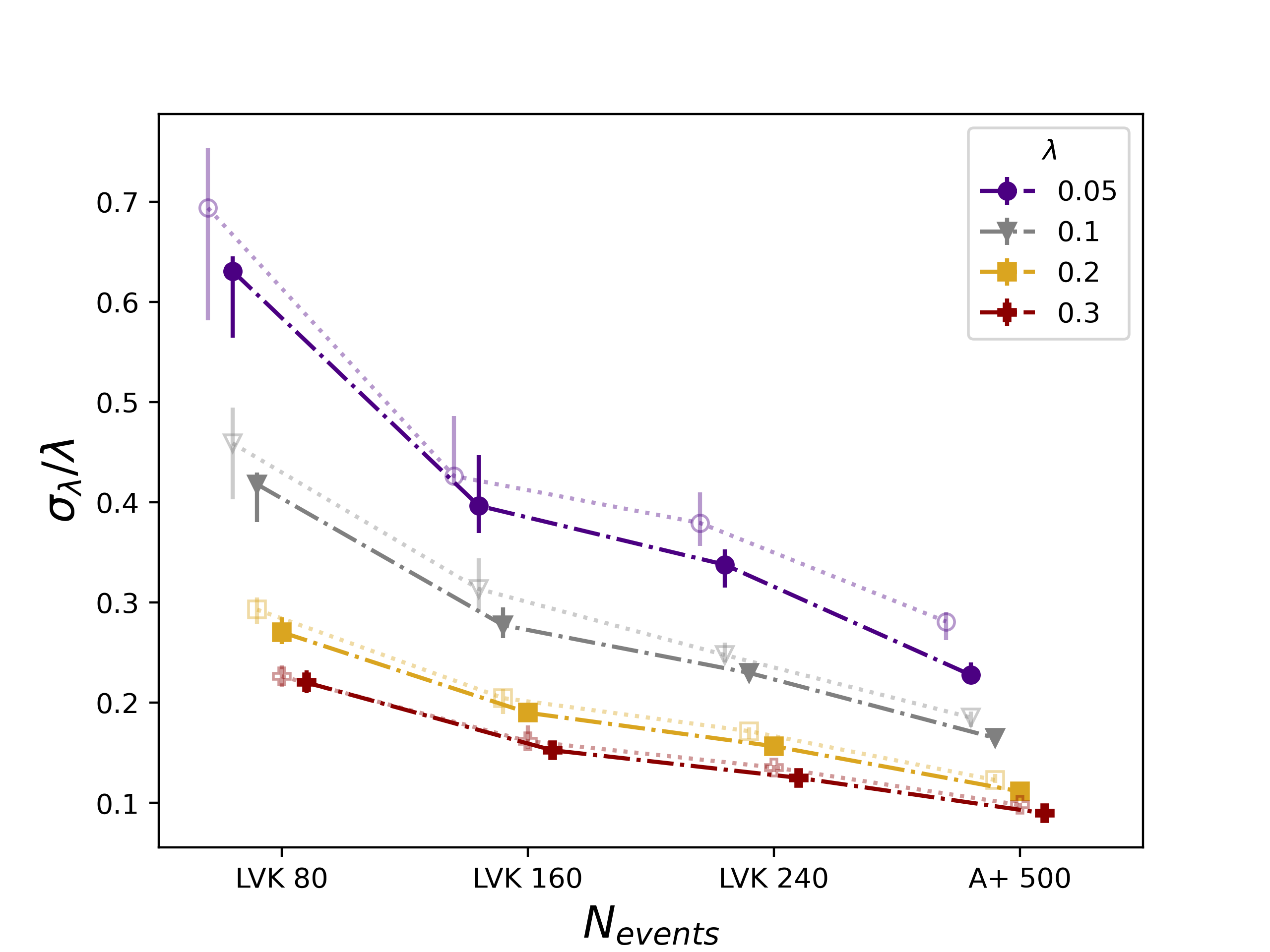}
\includegraphics[width=0.49\textwidth]{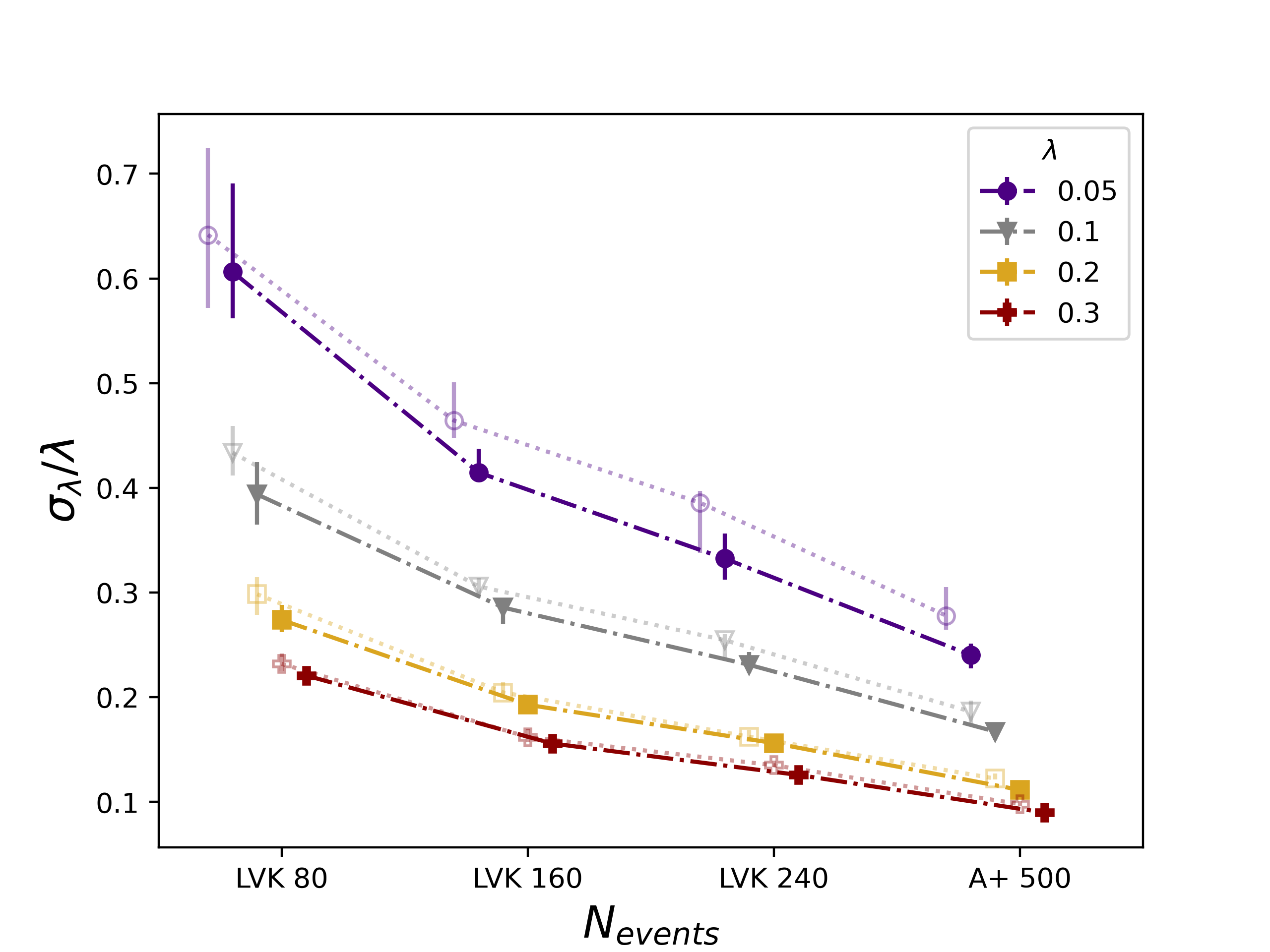}
\caption{Precision on the fraction of BBHs inducing a AGN flare, $\lambda$, as function of events in LVK (similar to the expected O4 sensitivity) and LVK A+ (similar to O5) for different true values $\lambda$.  The results in the top (bottom) panel assume a flat prior over $\Omega_m$ within $[0,1.0]$ ($[0.25,0.35]$). The number density of AGN is $n_{\rm AGN}=10^{-4.75}$  Mpc$^{-3}$ for the filled symbols and $10^{-4.0}$ Mpc$^{-3}$ for the empty symbols.  }\label{fig:lambda}
\end{figure}

In Figure \ref{fig:lambda} we show the constraints we recover on $\lambda$. While these are not the main result of this manuscript, it is interesting to note how $\lesssim 10-30\%$ precision measurements are possible for all cases explored here. These constraints will have important consequences in understanding the interplay between different BBH formation channels.

\subsection{AGN catalog case}

We also apply our formalism to the case in which no follow-up is performed, therefore only an AGN catalog without flare information is used. The recovered  precision on $H_0$ and $\lambda_{\rm AGN}$ after $1$ year ($80$ events) of LVK considering all AGNs in the GW area for $n_{\rm AGN}=10^{-4.75}$ Mpc$^{-3}$ and $\lambda_{\rm AGN}$ in the range of $[0.3,0.8]$, is summarized in Table \ref{tab:noflare}. In this scenario $\lambda_{\rm AGN}$, differently from $\lambda$, represents the fraction of all BBHs hosted in AGNs, and thus $\lambda_{\rm AGN} \geq \lambda$. The range of $\lambda_{\rm AGN}$ was chosen to follow the constraints on the BBH AGN channel from \cite{ford_mcKernan_21_rate}. We select the top $50\%$ events in terms of comoving volume sky map localization, as that is a measure of our background contamination, and only choose the events that will provide most of the constraining power while reducing the computing time needed (which is much larger in the non-flare case compared to the flare case, given the larger contamination). Although the results show a worse precision level of $H_0$ compared to the flare case for a specific value of $\lambda$, as expected, they still result in an interesting  precision at the $\sim 5-10\%$ level. Similarly to the AGN flares case, we consider the case where only BBHs on the high-mass end of the merging BH mass distribution merge in AGN disks. In that case, we find that the precision is significantly worsened, going from $5-10\%$ precision to $\sim 11-27\%$. This is expected as the most massive mergers are more likely to be at larger distances and encompass larger localization volumes. Compared to the flares case, the contamination from background AGNs plays a more important role given that there is a difference of several orders of magnitudes, hence the worsening due to the larger volumes is more pronounced in the AGN catalog case.



\section{Discussion}\label{sec:discussion}

In this section, we discuss a few important aspects of the standard siren method discussed here. First we consider the GW selection effects, then we discuss the case of BBHs in AGNs preferentially occupying the most massive section of the BBH mass spectrum as well as  caveats of the method, and finally move on to comparing to other standard siren methods. At last, we discuss some specific aspects of the case in which only AGN catalogs are avaialble instead of AGN flares.









\subsection{Selection effects}

For the flaring AGN case, we use the top $95\%$ events based on their comoving volume localization. This cut avoids all events with localization of thousands of squared degrees. This choice is reasonable for two reasons. First, the largest localization volumes have the largest number of background contaminants, hence will provide the least constraining power to the parameters of interest. Second, given their large typical sky localization, they are unlikely to prompt follow-up observations, or be serendipitously observed by sky survey over the majority of probability of their sky localization. 

For the case where we take into account all AGNs, i.e. without considering the flares, we impose a more stringent cut on the localization volume. We find this to be important because in the current formalism we are assuming a certain number density of AGNs, and then assume that same value in our posterior computation. In other words, we perfectly know the expectation value of the number density of AGNs (typically down to some luminosity). As a result, perfectly knowing the number of expected AGNs within a certain comoving volume carries information about the cosmology even without having any AGN associated to the GW events (i.e. from the background AGNs only, even for input $\lambda=0$, effectively without a standard siren measurement). This effect has a growing impact as the background density increases, and only becomes relevant for the static AGN catalog case, as the numbers we consider for the flares are too sparse. Ideally, one would leave the AGN number density free to vary in this case, with some reasonable prior to overcome this issue and present a more realistic scenario. We test this possibility for $\lambda_{\rm AGN}=0$, and find, as expected, that it did not carry over the extra volume constraining power in the cosmological constraints. However, since it is computationally challenging to run on a large set of cases (several runs for different $\lambda_{\rm AGN}$ values), we only constrain ourselves to the lowest background cases (i.e. lowest comoving volumes) where we have tested that the $\lambda_{\rm AGN}=0$ case does not provide cosmological parameter constraints that are more constraining than the priors.

A systematic bias on $H_0$ may arise from selection effects since a specific sample of detected GW  events is considered. First, to define an event detection, e.g. an SNR cut is applied. GW selection effects from this type of cuts affect a standard siren $H_0$ posterior approximately by a factor of $\propto H_0^{-3}$ \cite{chen17}, or, more generically if one considers other cosmological parameters, by a volume term. Because $H_0$ is the cosmological parameter we recover more precisely, we focus on this parameter. Since this study is focused on the precision, we disregard the influence of this selection bias in our analysis. In addition, as our formalism differs from traditional dark siren methods, our selection function includes a cosmology dependence in the $\mu_i$ term in Eq. \ref{eq:selection}. A study of the impact of these selection effects on future, more precise measurements will be presented in future work. It is also worth noting that a dark standard siren $H_0$ posterior, and specifically the selection effects term, depend on the black hole mass distribution and merger rate redshift evolution, hence making assumptions about these (e.g. fixing the shape and parameters of the mass distribution) when computing the posterior, may introduce biases in the recovered $H_0$ \cite{LVK21_StS}. On the other hand this effect diminishes as the redshift information from the EM observations (galaxy catalog for the dark siren case, AGN observations for this work) become increasingly informative \cite{LVK21_StS}. In the cases considered here, we are in a regime where the EM observations from AGNs are informative, hence it is possible that this effect does not have a significant impact given the level of statistical precision we obtain. Future work should focus on the impact of the mass distribution assumptions on the BBH-AGN formalism used here, as well as considering how to use AGN hosts in concert with spectral siren analyses as currently proposed with a generic host galaxy population \cite{2023PhRvD.108d2002M,Borghi:2023opd}.

The second selection criteria we apply is on the comoving volume. We make several tests on LVK 1 year simulations considering limited volumes (top $90\%$, $95\%$, $99\%$ events based on volume), and without any preselection for the case of the lowest considered value of $\lambda=0.05$ and highest number of background AGNs scenario with $\alpha=10^{-3}$ and $n_{\rm AGN}=10^{-4.5}$ Mpc$^{-3}$ and we find no significant change in the forecasts. More broadly, in our entire set of simulations we do not observe biases in the recovered posteriors when comparing to the input values (e.g. Figure \ref{fig:H0_w}, \ref{fig:corner}), and hence we claim that the impact of the selection effects should be smaller than the level of precision found here. On the other hand, the GW selection effects are not expected to have an impact on the recovery of $\lambda$, because, unlike the cosmology parameters, the aforementioned selection effects would not, in general, have a dependence on this parameter.

In this work, we have not taken into account EM selection effects, although \cite{Palmese_AGN} show how these can be taken into account in the present formalism. Unlike the GW selection effects, the EM selection effects are expected to have a more significant impact on the recovery of $\lambda$ rather than on the cosmology: failing to detect fainter AGNs or flares and not accounting for it, may result in recovering values of $\lambda$ that are biased low compared to the truth value, but it is not expected to affect the recovered cosmology precision as the non-detection of a flare will not have constraining power on the cosmology.
Since most of the events considered are at redshift $z\lesssim 0.6$ (see Fig. \ref{fig:sims}), it is also reasonable to expect that current and upcoming EM facilities (e.g. Euclid, Rubin Observatory, DESI) will provide complete catalogs of AGNs at these redshifts.

\begin{figure}[h]%
\centering
\includegraphics[width=0.49\textwidth]{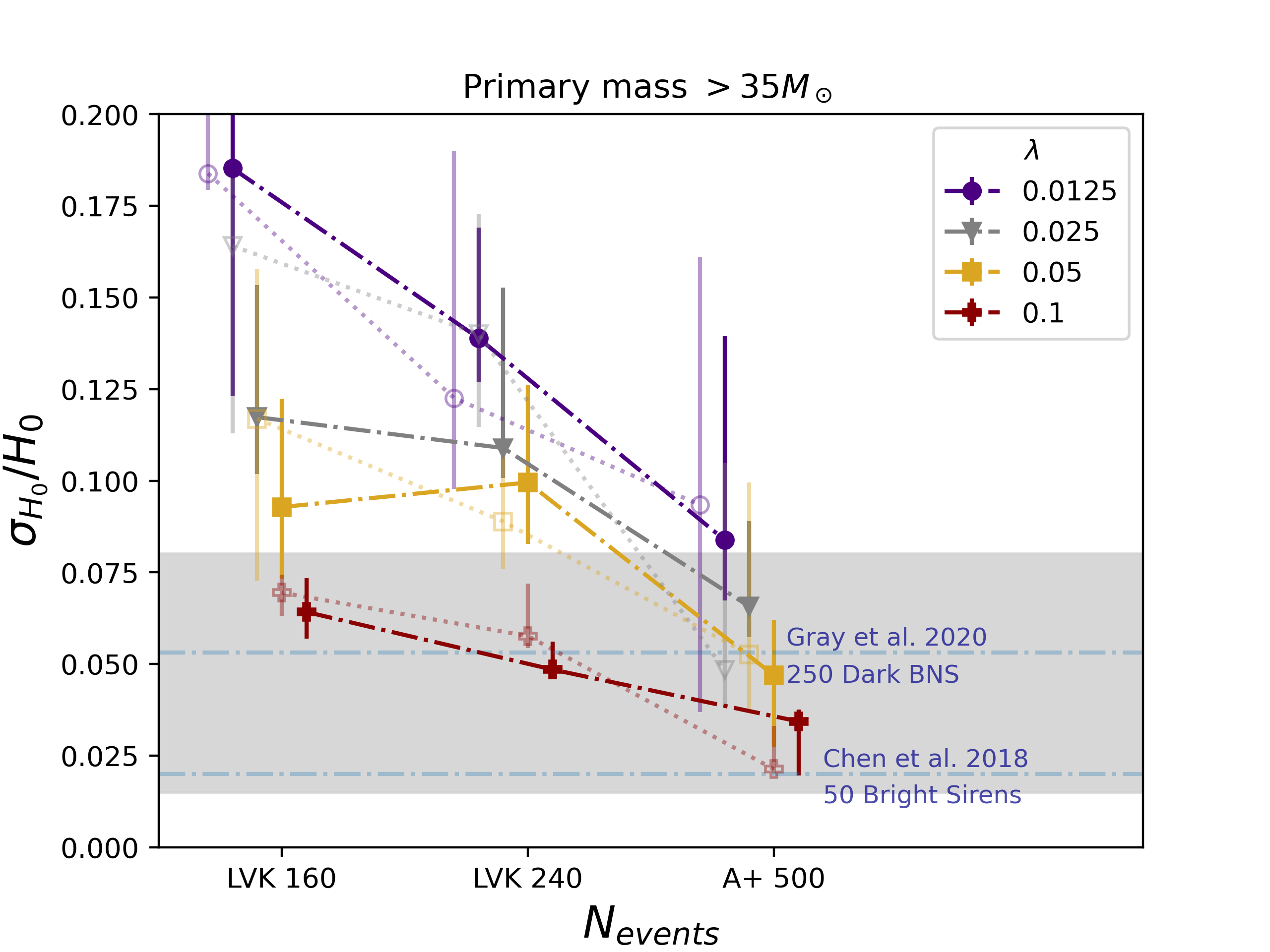}
\caption{Hubble constant precision as a function of  number of events from LVK design and LVK A+ for a scenario in which only BBH mergers with higher primary masses produce AGN flares. The upper dot-dashed horizontal line presents the expected $H_0$ precision from $250$ Dark Siren using BNSs from \cite{gray2020}, considering a $50\%$ complete galaxy catalog and unweighted galaxies. The lower dot-dashed line presents the forecast from \cite{chen17} after $\sim 50$ BNS bright sirens. The shaded region shows the uncertainties over $100$ realizations expected for a golden dark siren where we can identify the host from HLV+ with spectroscopic redshift. The number density AGN is $n_{\rm AGN}=10^{-4.75}$  Mpc$^{-3}$ for the filled symbols and $10^{-4}$ Mpc$^{-3}$ for the empty symbols. These results consider a flat prior on $\Omega_m \in [0.25,0.35]$.\label{fig:H0_precision3}}
\end{figure}

\subsection{BBH in AGNs as high mass mergers}

The findings of \cite{Gayathri_2021} suggest that the AGN channel is more likely to produce a higher fraction of the higher mass BBHs observed in the black hole mass distribution, than those produced below the $30-35 ~M_\odot$ peak \cite{GWTC_3,LVK:2021duu,Farah_2023}. This population of mergers is typically expected to occur at a greater distance and have less precise localization than the average detected BBH. On the other hand, the results shown in Figure \ref{fig:H0_precision} assume that the BBHs inducing AGN flares are a random subsample of the BBH population. To take this into account, we conduct two tests. First, we assume a scenario in which BBH mergers only occur in AGN disks if the primary mass $m_1$ is beyond the $\sim 30-35 ~M_\odot$ peak observed in the mass distribution. We assume $m_1> 35 ~M_\odot$ for the BBHs in AGNs for simplicity, as it is reasonable to believe that the peak may originate through a formation channel different from the AGN one, and this cut is justified considering that \cite{Farah_2023} finds a slightly lower value of the peak mass than previously found for GWTC-3. We present the results for this case in Figure \ref{fig:H0_precision3}. We find that the precision worsens, as expected, compared to our previous findings, although only slightly when comparing results for the same values of $\lambda$. This is due to the fact that the number of events occurring in AGNs, hence $\lambda$, is diminished (even if we assume that 100\% BBHs with $m_1> 35 ~M_\odot$ produce a flare, we cannot reach the $\lambda=$0.2 or 0.3 cases in Figure \ref{fig:H0_precision}), and because these massive mergers will typically have worse localizations and distance precision. Comparing e.g. the $\lambda=0.1$ case of Figure \ref{fig:H0_precision} with that of Figure \ref{fig:H0_precision3}, we find slightly larger scatter in the $H_0$ precision in the latter case, as even a single nearby event detection could be very loud and provide better distance measurements than lower mass objects at a similar distance.
For the second test, we also consider lower mass BBHs. If we consider that the higher mass BH in the mass distributions are formed hierarchically in the AGN disk, then it is also reasonable to assume that for each BBH with $m_1> 35 ~M_\odot$, there is at least another BBH of first generation black holes below that value that formed the second generation BH in the $m_1> 35 ~M_\odot$ binary, so we also sample for each high mass BBH in AGN a low mass BBH in AGN with primary below the peak. The results in this case become extremely similar to these in Figure \ref{fig:H0_precision} for the same values of $\lambda$, so we do not show them in the figure. We conclude that a 2-10\% precision in $H_0$ is also possible with higher mass BBHs merging in AGNs with a A+ sensitivity, assuming that $\lambda>1\%$.

\subsection{Caveats}

A caveat of the proposed method is that its performance is closely related to the fraction of BBH merging in AGN disks. GW population studies \cite{zevin2021} show that multiple BBH formation channels are likely at play, with dynamical formation being a viable scenario, but the contribution of each channel is yet to be well constrained. It is also worth noticing that our method, at least for the AGN flare case, is highly dependent upon the detectability of EM counterparts of BBHs inducing AGN flares, which is a subfraction of the BBHs hosted in AGN disks. The detectability  is closely related to the  interaction of BBHs and the AGN disks proprieties and is being currently investigated. This hypothesis has theoretical modelling support \cite{mckernanRam,rodriguez2023optical, tagawa23,kimura2021,BartosRapid}, and promising candidates from these channels have been reported \cite{graham20,graham23}. 
For instance, \cite{graham23} identified several new candidate EM counterparts to BBH events during the LVK O3 campaign. According to the authors, $7$ EM counterparts were statistically associated to BBH events (with a coincidence probability of the flares with all the associated BBH of $p=1.90\times 10^{-3}$) from the total set of $83$ events considered on the third observing run (O3). Their findings suggest that the fraction of detectable flares from BBHs is around $\sim 0.07$ in O3 within the ZTF explored volume up to magnitude $g<20.5$, considering ZTF observes roughly half of the sky. 


Another caveat for the forecast presented here is the assumption that some fraction of the \emph{observed} BBH population occurs in AGN, where there is no correlation between this specific formation channel and the BBH properties. In reality, it is possible that the AGN formation channel dominates the higher mass end of the BH mass distribution, thanks to its ability to retain hierarchical mergers.

It is important to mention that the forecasts presented here (as well as in most of the other standard siren forecast works quoted) assume the LVK detectors at a sensitivity (roughly the design sensitivity) that was expected until mid 2022 for the upcoming observing run O4. In reality, the Virgo detector has not joined the O4 run, and it is unclear whether its expected sensitivity for O4 will be achieved during this run. KAGRA has joined the run for a limited time at a significantly lower BNS inspiral range than what assumed here (80 Mpc). These changes to the observing capabilities only became publicly clear after this work was close to finalization. Therefore, the forecasts presented here are optimistic if considered for the upcoming O4 run, but they are still useful for future runs.

\subsection{Comparison with other standard siren methods}

Next, we consider how competitive and complementary our method is when compared to several other standard siren constraints expected from current generation GW detectors. 

First we consider the constraints shown for reference by the horizontal lines in Figure \ref{fig:H0_precision}. We stress that this is not a realistic comparison in terms of upcoming results since the analysis of \cite{gray2020} is based on BNS simulations that assume an O2--like sensitivity (hence, the same event detected in e.g. O4 should have an improved signal-to-noise ratio, hence localization, compared to those simulations on average). Moreover, 250 BNSs are unlikely to be detected during O4 (and potentially not even in 1 year of O5). Note that \cite{chen17} provides less optimistic forecasts from dark standard sirens with BBH mergers, potentially only reaching a 10\% precision with all detectors considered here at design sensitivity, while a BNS dark siren analysis is expected to reach that 5\% within such scenario. On the other hand, \cite{Borhanian} show how some extremely well-localized dark sirens could provide a $\sim 2-8\%$ precision (shown as shaded region in Figure \ref{fig:H0_precision}). It is also worth mentioning that a precision of $\sim 2.5-3\%$ is feasible within about a year of A+ if $\lambda$ is on the high-end ($\lambda \sim 0.2-0.3$) of the range we consider.  
\cite{gray2020} consider BNS mergers dark sirens which are typically closer than BBHs and less sensitive to $\Omega_m$. For a spectroscopic galaxy catalog which is 50\% complete, they find a 5\% precision on $H_0$, which is a constraint that our method can also achieve within the next $\sim5$ years as long as $\lambda>0.1$. 
The study in \cite{chen17} relies primarily on bright standard sirens, and finds that to reach the $2\%$ precision level would take around $50$ bright sirens. Considering the median detections expected in \cite{petrov2022data} scenarios, these figures are likely to be reached during O5. On the other hand, considering that only one BNS EM counterpart has been detected so far and that follow-up campaigns might not succeed on detecting all EM counterparts, due to limited telescope time or more edge-on or high mass systems dominated by a fainter and/or red component \cite{bom2023designing}, the number of events with confirmed counterparts may be significantly less than the number of BNS or NSBH detections. 
The addition of another sample of candidate EM counterparts and potential standard sirens is therefore interesting to ensure that GW cosmology is able to weigh in the Hubble constant tension before the 2030s.
 Compared to other bright standard siren classes, namely BNS and NSBH, the observation of flares might not require disruptive Target-of-Opportunity (ToO) imaging or even ToO spectra for the present study. First, there is no need to require a direct association of a given flare to a specific GW event, although ideally spectroscopic confirmation of e.g. off-center events through spectroscopy of asymmetric broad lines \cite{mckernanRam} would be a smoking gun for association (hence becoming a bright standard siren, if the uniqueness of the counterpart can be established). Additionally, the flare time scales are expected to be of the orders of weeks to months for higher mass AGNs, as opposed to less than 1 week for kilonovae. Hence, AGN flare counterpart searches would rather require all sky monitoring such as that carried out by ZTF or Rubin Legacy Survey of Space and Time (LSST), and a small spectroscopic follow-up campaign to acquire redshifts of flares where not available. 
For instance, in our high background contaminant scenario, our simulations show that $\sim 95\%$ of events with $n_{\rm AGN}=10^{-4.5}$ Mpc$^{-3}$ and $\alpha_{\rm flare}=10^{-3}$ would have less than $\sim 7000$ AGNs in the GW area and therefore around $7$ background flares potentially associated to an event.

\cite{Borhanian}  requires golden events where one can identify the exact BBH host galaxy, assuming $100\%$ duty cycle over 2 years of LVK at A+ sensitivity, to reach a $\sim 2\%$ precision level on $H_0$. Under these circumstances, dark sirens act as bright sirens, and are therefore extremely powerful. In these special cases, no active follow-up is required for the purpose of dark standard siren cosmology, although complete imaging and spectroscopy of the high probability region will be fundamental. On the other hand, discovering an AGN flare in coincidence with such golden events would constitute a smoking gun for the GW-AGN association and a higher confidence standard siren measurement, given the low probability of chance coincidence of flares over the small volumes considered. 

\citep{Farr_2019} present a method that constrains the Hubble parameter using GW BBHs alone by taking advantage of a specific feature in the mass distribution of black holes. Their method is most constraining at $H(z=0.8)$ with a set of $\sim 10^3$ to $5 \times 10^3$ events from the LVK network and can be translated to a precision in $H_0$ to $\sim 6\%$ ($\sim 10^3$ events) and $ \sim 3\%$ ($\sim 5\times10^3$) level. The method presented in this work is competitive, considering a free $\Omega_m \in [0.0,1.0]$ we use $240$ events for LVK to reach $\sim 5\%$ level for $\lambda=0.1$. 
However, the method of \cite{Farr_2019} has the advantage of not requiring any GW follow-up. The expectation of the $H(z)$ analysis is that it will enable constraints on $w$ to a $19\%$ ($\sim 10^3$ events) and $12\%$ ($\sim 5\times10^3$ events) precision after 1 and 5 years of LVK A+ (O5) running at design sensitivity, respectively, after imposing a $1\%$ prior on $H_0$ and the \emph{Planck} 2018 constraint on $\Omega_m$. Additionally, \cite{spectral_sirens} find slightly less constraining forecasts on $H(z)$ with an updated BBH population. Although we present wider uncertainties on $w$ our method is still promising compared with these proposed analyses, and perhaps more importantly, it is complementary. While most of the constraining power for the spectral sirens typically comes from the redshift and mass regions where the highest statistics of detections are available, specifically for 2G detectors, the lower bound of the pair-instability mass gap \cite{spectral_sirens}, or potentially the $\sim 30~M_\odot$ peak, and distances close to the detection horizon, in our case the constraints are more powerful at lower distances where better localizations are more likely. It will eventually be interesting to combine the two methods.


\subsection{AGN catalog case}
 
 At last, we discuss the case of follow-up for the AGN catalog case, where the flares are not taken into account. Considering only the best localized events will limit the number of AGNs in the field for which spectroscopic redshifts are required to $\sim 50-70$ at maximum per event, and thus feasible within the reach of many current spectroscopic facilities for a dedicated campaign. Such events are also the most relevant events since they have a lower number of background contaminants and lower uncertainties in $d_L$. 
  The usage of the present methodology with the full set of AGNs in the field can be interpreted as a type of informed dark standard siren method, which has the advantage of not requiring full galaxy catalogs, but the disadvantage of not providing constraining power for events that do not occur in AGNs. Our forecasts show that a 5-10\% precision on $H_0$ could be achieved within 1 year of LVK at design sensitivity in the optimistic case that 30-80\% of BBHs occur in AGNs. In the future, it will be interesting to expand on this promising avenue by taking into account AGN clustering in the simulations, and by further speeding up the sampling to produce full ensembles of simulated scenarios.

We also briefly discuss the role of completeness of AGN catalogs. For what concerns the AGN flares follow-up, although complete AGN catalogs would facilitate galaxy-targeted searches, they are not necessary for the method. For wide field searches, such as those that can be carried out with ZTF and LSST, it is reasonable to search for nuclear transients regardless of existing catalogs. Since nuclear flares of the type and shape that we would search for are typically rare, if some of the flares do not have an AGN classification, one could take additional observations to clarify their nature.  Similarly, one may consider to observe the localization of the GW event if a BBH shows features that may hint to an AGN formation, to produce an AGN catalog down to the desired luminosities. With the upcoming completion of large spectroscopic surveys such as DESI \cite{desicollaboration2016desi}, and the start of new surveys \cite{schlegel2022spectroscopic}, it is possible that dedicated follow-up will not be needed as complete catalogs of Type I AGNs will be available.
For both the flare and the AGN catalog case, the formalism allows us to specify the selection function of AGNs and flares (see Eq. \ref{eq:selection}). For both cases, it will be interesting in the future to quantify which AGNs luminosities are likely to contribute significantly to the BBH merger rates, and apply the selection function for those (efforts in that direction already exist, e.g. \cite{Ford_LINERS}).

\section{Conclusions}

In this study, we present forecasts using a novel and competitive method to obtain cosmological constraints from multi-messenger observations of gravitational wave events. Our work is build upon the formalism proposed in \cite{Palmese_AGN}, where the perspective of using AGN flares to constrain cosmology is pointed out. From a set of simulations representing a possible sample of detections from the upcoming LVK observing runs, we obtain constraints for cosmological parameters within the $\Lambda$CDM and $w$CDM models using multiple realizations. 

We find interesting constraints on the Hubble constant are possible, down to $3-6\%$ precision for O4 and O5-like runs, assuming that at least $\sim10\%$ of BBHs occurs in AGN disks.  We reach more competitive constraints on $H_0$, as opposed to $\Omega_m$ or $w$, because most of the constraining power comes from the events localized within a smaller volume, typically the most nearby, at $z<0.2$. The constraints we find will of course only be competitive if a significant fraction of LVK BBH mergers occur in AGNs, which is yet to be confirmed. Assuming that the AGN origin of BBHs can be confirmed, future work on this subject should explore in more detail how selection effects affect standard siren constraints from the proposed method.



\acknowledgments

Antonella Palmese thanks Maya Fishbach, Saavik Ford, Matthew Graham, Barry McKernan, Saul Perlmutter for useful discussion, Leo Singer for help with the injections and sky map software and for making BAYESTAR public, and the LIGO collaboration for providing public access tools. Antonella Palmese acknowledges support for this work was provided by NASA through the NASA Hubble Fellowship grant HST-HF2-51488.001-A awarded by the Space Telescope Science Institute, which is operated by Association of Universities for Research in Astronomy, Inc., for NASA, under contract NAS5-26555. Clecio Bom acknowledges the financial support from CNPq (316072/2021-4) and from FAPERJ (grants 201.456/2022 and 210.330/2022) and the FINEP contract 01.22.0505.00 (ref. 1891/22). The authors made use of Sci-Mind servers machines developed by the CBPF AI LAB team and would like to thank A. Santos P. Russano and M. Portes de Albuquerque for all the support in infrastructure matters.

\bibliography{sn-bibliography}


\end{document}